\documentclass[journal]{IEEEtran}

\usepackage{subfig}
\usepackage{float}
\usepackage{algorithm}
\usepackage{algpseudocode}
\usepackage{graphicx}
\usepackage{amsmath,amssymb,amsfonts}
\usepackage{url}
\usepackage{hyperref}
\usepackage{balance}

\begin{document}

\title{Zero-disparity Distribution Synthesis: Fast Exact Calculation of Chi-Squared Statistic Distribution for Discrete Uniform Histograms}

\author{Nikola Bani{\'{c}} and Neven Elezovi{\'{c}} 
}

\markboth{Journal of \LaTeX\ Class Files, Vol. 14, No. 7, July 2024}
{Shell \MakeLowercase{\textit{et al.}}: Bare Demo of IEEEtran.cls for IEEE Journals}
\maketitle

\renewcommand\footnoterule{{\hrule height 0pt}}

\renewcommand\footnoterule{{\hrule height 0pt}}
\let\thefootnote\relax\footnotetext{
Manuscript received.

Nikola Bani{\'{c}} is with Gideon Brothers, 10000 Zagreb, Croatia. Neven Elezovi{\'{c}} is with Department of Applied Mathematics, Faculty of Electrical Engineering and Computing, University of Zagreb, 10000 Zagreb, Croatia (email: nbanic@gmail.com, neven@element.hr).

Digital Object Identifier
}

\begin{abstract}
Pearson's chi-squared test is widely used to assess the uniformity of discrete histograms, typically relying on a continuous chi-squared distribution to approximate the test statistic, since computing the exact distribution is computationally too costly. While effective in many cases, this approximation allegedly fails when expected bin counts are low or tail probabilities are needed. Here, Zero-disparity Distribution Synthesis is presented, a fast dynamic programming approach for computing the exact distribution, enabling detailed analysis of approximation errors. The results dispel some existing misunderstandings and also reveal subtle, but significant pitfalls in approximation that are only apparent with exact values. The Python source code is available at~\url{https://github.com/DiscreteTotalVariation/ChiSquared}.
\end{abstract}


\IEEEpeerreviewmaketitle

\newcommand{\E}[1]{\mathbb{E}\left[ #1 \right]}
\newcommand{\A}[1]{\lvert #1 \rvert}
\newcommand{\Var}[1]{\mathrm{Var}\left[ #1 \right]}
\newcommand{\V}[1]{\left\lVert #1 \right\rVert_{V}}
\newcommand{\Norm}[2]{\left\lVert #1 \right\rVert_{#2}}
\newcommand{\N}[2]{\mathcal{N}\left(#1,#2\right)}

\section{Introduction}
\label{sec:introduction}

\subsection{Pearson's chi-squared test}
\label{subsec:pearson}

\IEEEPARstart{P}{earson's} chi-squared test is a cornerstone of statistical analysis, primarily used to check if a significant discrepancy exists between observed frequencies and those expected under a given hypothesis. A fundamental application of this is the goodness-of-fit test, which evaluates how well a sample distribution conforms to a hypothesized one. A common and critical use case is testing for a discrete uniform distribution, where each category or bin in a histogram is expected to contain an equal number of observations. This is vital in numerous fields, including random number generator evaluation~\cite{knuth1997art,bassham2010sp}, quantization and dithering analysis~\cite{lipshitz1991quantization,kobus2024gaussian}, statistical modeling of signals and noise~\cite{li2017probability}, statistical testing analysis~\cite{bland2013baseline}, genetics~\cite{bailey1995statistical}, quality control~\cite{montgomery2020introduction}, sampling~\cite{vonta2008statistical}, etc.

The discrete uniform distribution is a fundamental concept in statistics. A comprehensive understanding of the exact behavior of the chi-squared statistic in this context provides a crucial benchmark and deeper insight into the properties of goodness-of-fit tests in general. Thus, it is a good starting point for more detailed research into other exact distributions.

Let $\mathbf{x}_n$ be a discrete histogram that divides a sample of size $N$ into $n$ bins where $x_i$ is the count for the $i$-th bin so that
\begin{equation}
\label{eq:sum}
\sum_{i=1}^{n}x_i=N.
\end{equation}
The chi-squared statistic~\cite{davis2016handbook} used in the Pearson's test is
\begin{equation}
    \label{eq:chi}
    \chi^2_{c}=\sum_{i=1}^{n}\frac{\left(O_i-E_i\right)^2}{E_i}
\end{equation}
where $c$ is the number of degrees of freedom, $O_i$ is the $i$-th observed value, and $E_i$ is the $i$-th expected value. For uniform histograms, every $E_i$ is the same, and $c$ is $n-1$ since the $n$-th value can be deduced from $N$ and Eq.~\eqref{eq:sum}, which thus gives
\begin{equation}
    \label{eq:chi2}
    \chi^2_{n-1}=\sum_{i=1}^{n}\frac{\left(x_i-\frac{N}{n}\right)^2}{\frac{N}{n}}=\frac{n}{N}\sum_{i=1}^{n}x_i^2-N=\frac{n}{N}s-N.
\end{equation}
where $s$ is the sum of the squared bin values. The chi-squared statistic can here be computed directly from $s$ by scaling and shifting as shown in Eq.~\eqref{eq:chi2}, and the converse also holds:
\begin{equation}
    \label{eq:s}
    s=\left(\chi^2_{n-1}+N\right)\frac{N}{n}=\sum_{i=1}^{n}x_i^2.
\end{equation}
This integer form, $s$, can be useful in simplifying and speeding up calculations, and it will be used later here in this paper.

Once the chi-squared statistic from Eq.~\eqref{eq:chi2} is computed for a given sample, its (exact or approximate) distribution is used to calculate the $p$-value, which then informs the decision to accept or reject the null hypothesis, and to conclude the test.

\subsection{The approximation problem}
\label{subsec:problem}

The traditional Pearson's chi-squared test is an asymptotic test~\cite{dasgupta2008asymptotic}. Its reliability is predicated on the statistic in Eq.~\eqref{eq:chi2} following a continuous chi-squared distribution, an approximation that is valid only when sample sizes are sufficiently large. Even though this approximation is generally considered acceptable and robust~\cite{larntz1978small}, a widely accepted guideline is that all expected frequencies within the histogram bins should be at least $5$ ~\cite{cochran1952chi2,cochran1954some,yarnold1970minimum,andres2000minimum,decoursey2003statistics,moore2009introduction,bluman2014elementary,siegel2016practical,lock2020statistics}. When this condition is not satisfied, which is frequently the case with small sample sizes or histograms with numerous bins, i.e., sparse data, the standard test can yield inaccurate $p$-values. This may lead to erroneous conclusions, such as the incorrect rejection of a true null hypothesis (a Type I error) or the failure to detect a genuine deviation from the hypothesized distribution (a Type II error). For discrete uniform histograms, particularly in data-limited scenarios, expected frequencies can fall below the threshold of $5$, thus compromising the reliability of the test, if this threshold is even to be observed~\cite{agresti2013categorical}.

To overcome the limitations of the asymptotic approach in problematic cases of discrete data, it would be essential to determine the exact distribution of the chi-squared statistic. This would involve computing the precise probability of every possible value of the chi-squared statistic under the null hypothesis, rather than depending on an approximation or its corrections~\cite{yates1934contingency,conover1974some} whose behavior is not fully known.

Namely, one of the problem with approximations is that the exact magnitude of their errors is rarely quantified due to the computational cost of deriving the true distribution of the chi-squared statistic for specific $N$ and $n$. Hence, knowledge of the exact distribution is not only essential in fields that demand it, but also instrumental in more accurately assessing the error associated with commonly used approximation techniques.

As a matter of fact, it can be shown that, in the case of uniform distribution, a faster exact calculation of chi-squared statistic distribution is possible, as presented in this paper.

\subsection{Contribution}
\label{subsec:contribution}

In this paper, three main contributions are presented. First, a dynamic programming~\cite{bellman2015applied} method for the efficient computation of the exact distribution of the chi-squared statistic under the discrete uniform distribution is proposed. Second, a numerical analysis of approximation errors is conducted using the exact results, revealing that some commonly accepted rules regarding the relationship between $N$ and $n$ may not always be optimal. Third, an issue with approximated $p$-values for higher values of the chi-squared statistic is identified and examined in greater detail, with implications shown on real-life examples.

The source code with comments and pre-calculated data is given at~\url{https://github.com/DiscreteTotalVariation/ChiSquared}.

The paper is structured as follows: Section~\ref{sec:related} describes the related work on calculating the exact distribution of the chi-squared statistic, Section~\ref{sec:proposed} proposes a significantly faster exact calculation, experimental results with statistical tests are presented in Section~\ref{sec:results}, and Section~\ref{sec:conclusions} concludes the paper.

\section{Related work}
\label{sec:related}

An alternative to using $p$-values based on the chi-squared distribution for uniform discrete histograms that is often more accurate and relatively easy to compute is to perform Monte Carlo simulations~\cite{perkins2011chi2}. However, if the exact distribution is required to compute exact $p$-values, only a few approaches have been published to date. The standard method involves enumerating all possible histograms and checking them, as described in Section~\ref{subsec:naive}, though some optimized methods have also been developed, as discussed in Section~\ref{subsec:optimized}.

\subsection{Naive exact distribution calculation}
\label{subsec:naive}

Exact distribution calculation for uniformity tests was described already in~\cite{siegel1979noncentral} where statistical tests for uniformity over discrete distributions have been examined, focusing on the behavior of the chi-squared test under the null hypothesis. The importance of understanding the exact, discrete distribution of the chi-squared statistic was emphasized, especially in small-sample settings, highlighting discrepancies with its asymptotic chi-squared approximation. It was shown that for uniform multinomial data, the exact distribution of the test statistic can be derived by enumerating all possible histograms, though this becomes computationally infeasible as sample size or the number of bins increases if standard methods are used.

The standard method for computing the exact distribution of a statistic, including the chi-squared statistic, involves evaluating every possible histogram $\mathbf{x}_n$ of size $N$ that satisfies the desired condition~\cite{bishop2007discrete,agresti2013categorical,indrayan2017medical}. For a given histogram, its chi-squared statistic value is calculated, and the $p$-value is obtained by summing the probabilities of all histograms with chi-squared statistic values greater than or equal to it.

When distributing $N$ observations into $n$ bins, there are $n^N$ possible sequences of assignments, but many result in the same histogram. The number of unique histograms is given by the \textit{stars and bars} formula as $\binom{N + n - 1}{n - 1}$. Under a uniform multinomial distribution, the probability of a particular configuration $(x_1, x_2, \dots, x_n)$ is $\frac{N!}{x_1!x_2! \dots x_n}\left(\frac{1}{n}\right)^N$. Since every distinct configuration must be evaluated, the computational complexity is $O\left(\binom{N + n - 1}{n - 1}\right)$, i.e., $O\left(N^{n-1}\right)$ if $n$ is regarded as a constant, rendering this approach infeasible for $n>5$.

\subsection{Optimized exact distribution calculation}
\label{subsec:optimized}

Only a few recent papers share similar goals with this work. The most useful and significant of them are discussed below.

A somewhat related paper~\cite{bonetti2019computing}
presents efficient methods to compute the exact finite-sample distributions of order-based statistics; specifically the maximum, minimum, range, and sum of the J largest cell counts from a uniform multinomial distribution. While these statistics are useful for goodness-of-fit testing and therefore somewhat related, the paper does not compute the exact distribution of the chi-squared statistic itself, which involves a distinct and more complex enumeration of all possible histograms exceeding a given chi-squared value.

A significant speedup over naive enumeration has been presented in~\cite{resin2023simple} where an efficient algorithm is introduced to compute exact $p$-values for goodness-of-fit tests under the multinomial distribution. Rather than enumerating the entire sample space, the algorithm incrementally explores outcome vectors within increasing $\ell_1$-distance from the expected count vector under the null hypothesis. This search strategy leverages discrete convexity to focus computational effort on the most relevant parts of the space. The method supports common test statistics such as the chi-squared statistic.  A key improvement over naive enumeration is the complexity of the approach: rather than the full enumeration's $O\left(N^{n-1}\right)$ complexity, the algorithm only examines an acceptance region achieving complexity $O\left(N^{(n-1)/2}\right)$. However, this complexity still increases exponentially with $n$, making it impractical for large $n$ and $N$.

While the described approaches are definitely an improvement over the naive approach, they are still impractical for any larger values of $N$ and $n$, hence a better solution is required.

\section{Faster exact distribution calculation}
\label{sec:proposed}

There are $n^N$ equally likely bin assignment sequences under a uniform distribution. However, the number of distinct chi-squared statistic values is much smaller, fewer than $N^2$, due to the discrete nature of histogram bins. Rather than computing the chi-squared statistic for every possible histogram, it is more efficient to track how the statistic's value counts change as bins are added to an existing histogram with known counts.

A general idea for this approach is given in Section~\ref{subsec:idea}, the calculation specifics are in Section~\ref{subsec:specifics}, the complexity of this approach is presented in Section~\ref{subsec:complexity}, Sections~\ref{subsec:probabilities},~\ref{subsec:reusing}, and~\ref{subsec:correct} explain how to correctly calculate the $p$-values for individual chi-squared statistic values, Section~\ref{subsec:algorithm} goes over some implementation details that can reduce the computation cost, Section~\ref{subsec:other} discusses using other statistics, Section~\ref{subsec:nonuniform} comments on extending the described procedure to non-uniform distributions, and Section~\ref{subsec:name} names the whole procedure for simplicity.

\subsection{The general idea}
\label{subsec:idea}

The chi-squared statistic value counts are first computed for a histogram with only a single bin and this is done for all sample sizes from $1$ to $N$. Then, bins are added one at a time, and at each step, the statistic value counts are recalculated for all sample sizes and their possible bin arrangements. This process continues until finally the histogram has $n$ bins, at which point the required distribution counts are obtained.

\subsection{Calculation specifics}
\label{subsec:specifics}

Let $C_{N, n}(i, M, s)$ be the count of bin assignment sequences where a sample of size $M$, itself a subsample of a sample of size $N$, fills $i$ bins so that the resulting histogram has a sum of squared bin values $s$, where $1 \leq i \leq n$, $0 \leq M \leq N$, and $0 \leq s \leq N^2$. Knowing $C_{N, n}(1, M, s)$ allows recursive computation of $C_{N, n}(2, M, s)$, up to $C_{N, n}(n, N, s)$, which are used for exact probability calculations in Section~\ref{subsec:probabilities}.


\subsubsection{Single bin}
\label{subsubsec:single}

For $i=1$, a histogram with a single bin, the sum of squared bins is fixed for each $M$ and given as
\begin{equation}
    \label{eq:single}
    s'=M^2.
\end{equation}
There are $\binom{N}{M}$ ways to assign $M$ samples to this bin, so $C_{N, n}(1, M, s')=\binom{N}{M}$, and $C_{N, n}(1, M, s)=0$ for all $s \ne s'$.

\subsubsection{Added bins}
\label{subsubsec:added}

Regarding $C_{N, n}(i, M, s)$, a histogram with $i > 1$ bins and sample size $M$ is created by adding a bin with value $m$ to an existing histogram with $i-1$ bins and sample size $M-m$. If the sum of squared bins of the histogram with $i-1$ bins is $s'$, then for the new histogram with $i$ bins it is
\begin{equation}
    \label{eq:new_s}
    s=s'+m^2.
\end{equation}
The $i$-th bin is filled to $m$ by choosing from $N-(M-m)$, which can be done in $\binom{N-M+m}{m}$ ways. This can be done for any $m$ such that $0 \leq m\leq M$, and therefore $C_{N, n}(i, M, s)$ is
\begin{equation}
    \label{eq:dp}
    \begin{gathered}
    C_{N, n}(i, M, s) \\
    =\sum_{m=0}^{M}\binom{N-M+m}{m}\cdot C_{N, n}(i-1, M-m, s-m^2).
    \end{gathered}
\end{equation}

\subsection{Complexity}
\label{subsec:complexity}

Calculating all values of $C_{N, n}(i, M, s)$ involves $n$ values of $i$, $N+1$ of $M$, and roughly $N^2$ values of $s$, resulting in a computational complexity of $O(nN^4)$. The memory complexity is $O(nN^3)$, but it can be reduced to $O(N^3)$ by storing counts only for $i$ and $i-1$, i.e., current and previous counts. Additionally, while Eq.\ref{eq:dp} is correct, a slightly different calculation order, detailed in Section~\ref{subsec:algorithm}, is better for speed.

If $C_{N, n}(i, M, s)$ is only needed for a specific statistic value $t$, computations for $s>t$ can be skipped to reduce complexity.

\begin{algorithm}
\caption{Getting the exact chi-squared statistic distribution}
\label{alg:p}
\hspace*{\algorithmicindent}\textbf{Input:} $N$, $n$ \\
\hspace*{\algorithmicindent}\textbf{Output:} $p_{N, n}^{(0)}$, $p_{N, n}^{(1)}$, $\ldots$, $p_{N, n}^{(N^2)}$
\begin{algorithmic}[1]
\State cc $\gets$ Zeros($N+1$, $N^2$) \Comment{current counts}
\For{$i\in\{1, 2, \ldots, N\}$} \Comment{initialize the first bin's counts}
    \State cc$(i, i^2)\gets \binom{N}{i}$
\EndFor
\For{$b\in\{2, \ldots, n\}$}
    \State pc $\gets$ cc \Comment{earlier current counts become previous}
    \State cc $\gets$ Zeros($N+1$, $N^2$) \Comment{clear the values}
    \For{ps $\in\{0, \ldots, N\}$} \Comment{previous sum}
        \State r $\gets$ $N$ $-$ ps \Comment{remaining values}
        \For{p\_chi $\in\{0, \ldots, N^2\}$} \Comment{previous sum of squared bin values}
            \State prev $\gets$ pc(ps, p\_chi)
            \If{prev$>0$}\Comment{important for speed}
                  \For{nl $\in\{0, \ldots, $r$\}$} \Comment{last bin value}
                    \State ns $\gets$ ps $+$ nl \Comment{new sum}
                    \State n\_chi $\gets$ p\_chi $+nl^2$
                    \State
                    \parbox[t]{\dimexpr\linewidth-\algorithmicindent}{cc(ns, n\_chi) $\gets$ \\ cc(ns, n\_chi) $+$ prev $\cdot$ $\binom{\text{r}}{\text{nl}}$
                    }
                \EndFor
            \EndIf
        \EndFor
    \EndFor
\EndFor
\For{$i\in\{1, 2, \ldots, N^2\}$}
    \State $p_{N, n}^{(i)}=\frac{cc(N, i)}{n^N}$
\EndFor
\end{algorithmic}
\end{algorithm}

\begin{figure*}[htb]
    \centering
    
	\subfloat[]{
	\includegraphics[width=0.45\linewidth]{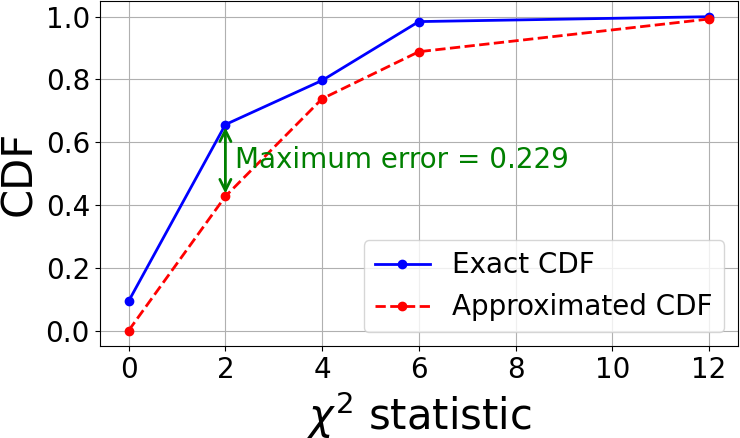}
	\label{fig:N_4_n_4}
	}%
    \subfloat[]{
	\includegraphics[width=0.45\linewidth]{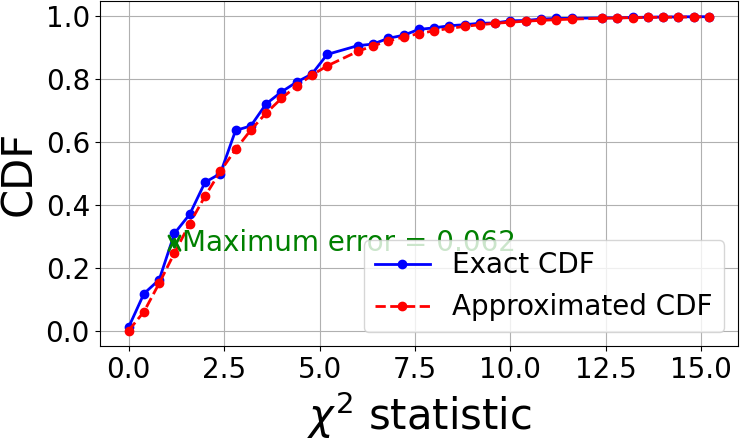}
	\label{fig:N_20_n_4}
	}\\
	
    \subfloat[]{
	\includegraphics[width=0.45\linewidth]{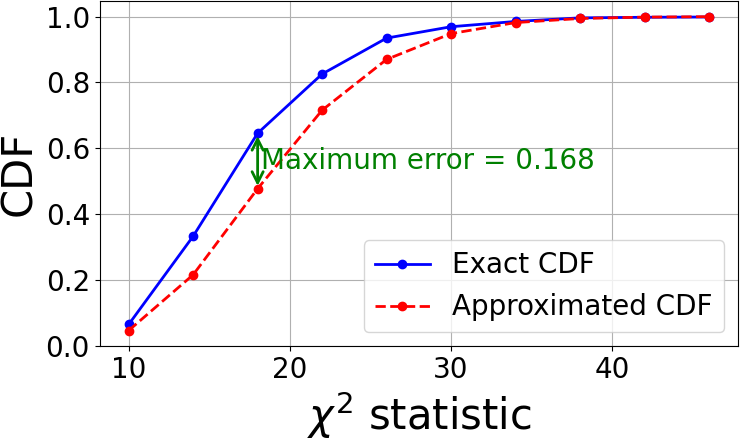}
	\label{fig:N_10_n_20}
	}%
    \subfloat[]{
	\includegraphics[width=0.45\linewidth]{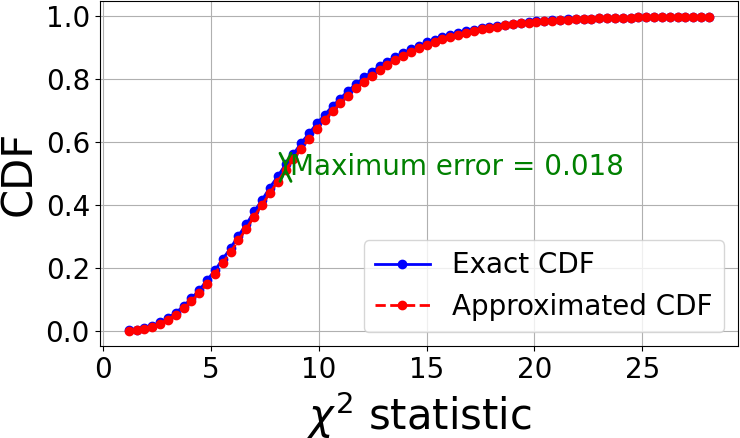}
	\label{fig:N_55_n_10}
	}\\


	\subfloat[]{
	\includegraphics[width=0.45\linewidth]{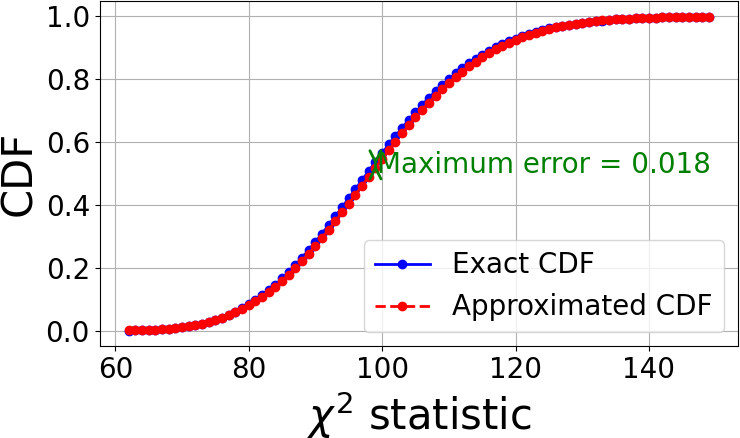}
	\label{fig:N_200_n_100}
	}%
	\subfloat[]{
	\includegraphics[width=0.45\linewidth]{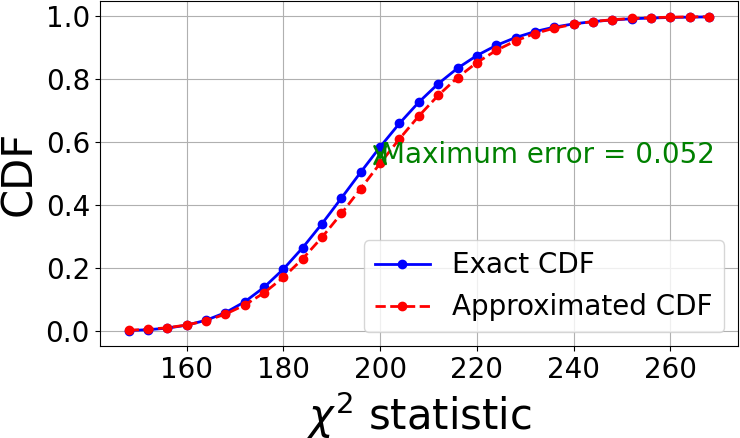}
	\label{fig:N_100_n_200}
	}\\

    \caption{Comparing the exact distribution of the chi-squared statistic for uniform histograms to the chi-squared distribution approximation when a)~$N=4$ and $n=4$, b)~$N=20$ and $n=4$, c)~$N=10$ and $n=20$, d)~$N=55$ and $n=10$, c)~$N=200$ and $n=100$, and d)~$N=100$ and $n=200$. With the exception of a), the tails are not displayed here due to their flatness.}
	\label{fig:comparing}
    
\end{figure*}

\subsection{The probabilities}
\label{subsec:probabilities}

A probabilistic interpretation of an observed chi-squared statistic value $\chi^2_{n-1}$ requires converting it into the corresponding sum of squared bin values $s$ in accordance with Eq.~\eqref{eq:s}. The exact probability of observing a sum of squared bin values $s$ from a uniform sample of size $N$ over $n$ bins is given as

\begin{equation}
    \label{eq:p}
    p_{N, n}^{(s)}=\frac{C_{N, n}(n, N, s)}{n^N}.
\end{equation}
The exact cumulative distribution function~(CDF) is given as
\begin{equation}
    \label{eq:cdf}
    P_{N, n}(s) = \sum_{\substack{t \leq s \\ t \in \mathbb{S}_{N,n}}} p_{N, n}^{(t)}.
\end{equation}
where $\mathbb{S}_{N,n}$ is the set of all values that the sum of squared bin values can take for histograms for given values of $N$ and $n$.

\subsection{Reusing the counts}
\label{subsec:reusing}

The counts $C_{N, n}(n, N, s)$ for various values of $s$ are used in Eq.~\eqref{eq:p} to calculate the probabilities of $s$ on histograms of uniform samples of size $N$ over $n$ bins. However, they can also be reused to calculate the probabilities of $s$ on histograms of uniform samples of size $M\leq N$ over $m\leq n$ bins:
\begin{equation}
    \label{eq:p_reused}
    p_{M, m}^{(s)}=\frac{C_{N,n}(m, M, s)}{m^M}\cdot \frac{1}{\binom{N}{M}}.
\end{equation}
The first term on the right side is as in Eq.~\eqref{eq:p}. The second term takes into account that the counts of $s$ for a subsample of size $M$ of a sample of size $N$ are added for $\binom{N}{M}$ subsamples.

\subsection{Correct $p$-value calculation}
\label{subsec:correct}

If the $p$-value for $s$ is calculated by approximating the exact distribution with the chi-squared distribution as usual like
\begin{equation}
    \label{eq:p_continuous}
    p=1-CDF_{\chi^2_{n-1}}\left(\frac{n}{N}s-N\right),
\end{equation}
this introduces an additional error since $s$ belongs to a discrete set $\mathbb{S}_{N,n}$ and $p_{N, n}^{(s)}>0$. To account for this, let $s^-$ be the largest value in $\mathbb{S}_{N,n}$ such that $s^- < s$. The exact $p$-value is
\begin{equation}
    \label{eq:p_discrete}
    p=1-P_{N,n}(s^-)=\sum_{\substack{t \geq s \\ t \in \mathbb{S}_{N,n}}} p_{N, n}^{(t)}.
\end{equation}

\begin{figure*}[htb]
    \centering
    
    \subfloat[]{
	\includegraphics[width=0.45\linewidth]{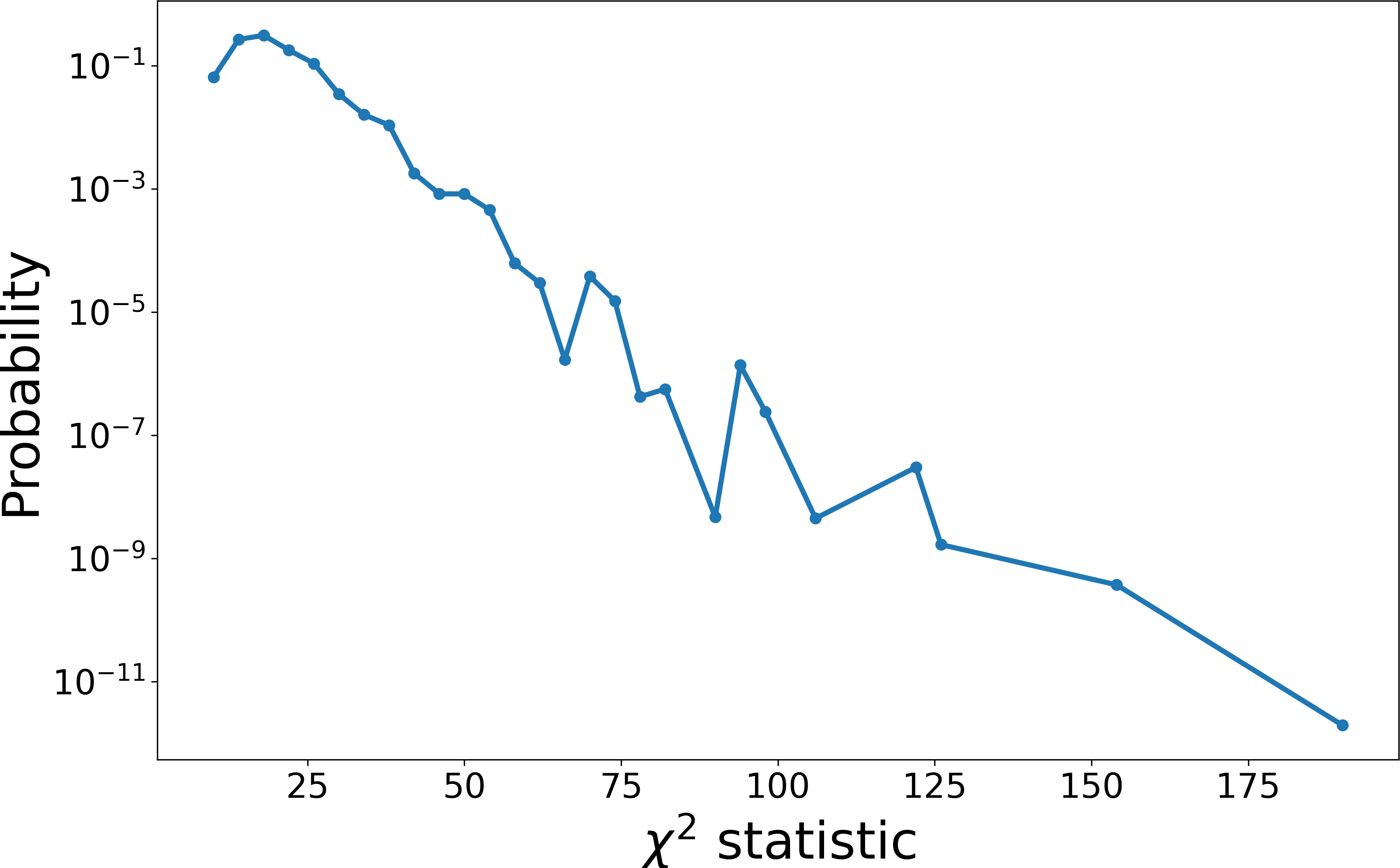}
	\label{fig:log_p_N_10_n_20}
	}%
    \subfloat[]{
	\includegraphics[width=0.45\linewidth]{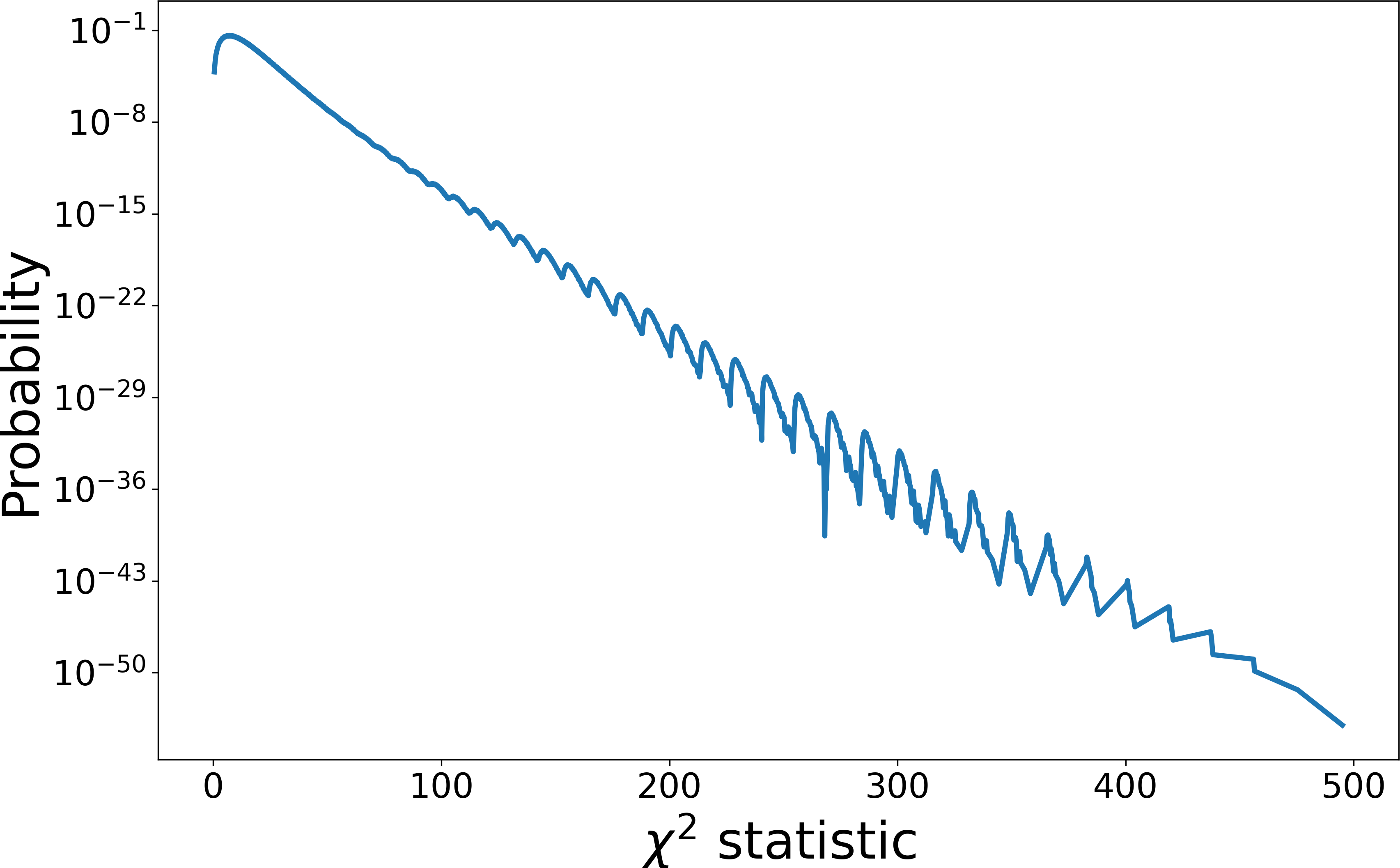}
	\label{fig:log_p_N_55_n_10}
	}\\

	\subfloat[]{
	\includegraphics[width=0.45\linewidth]{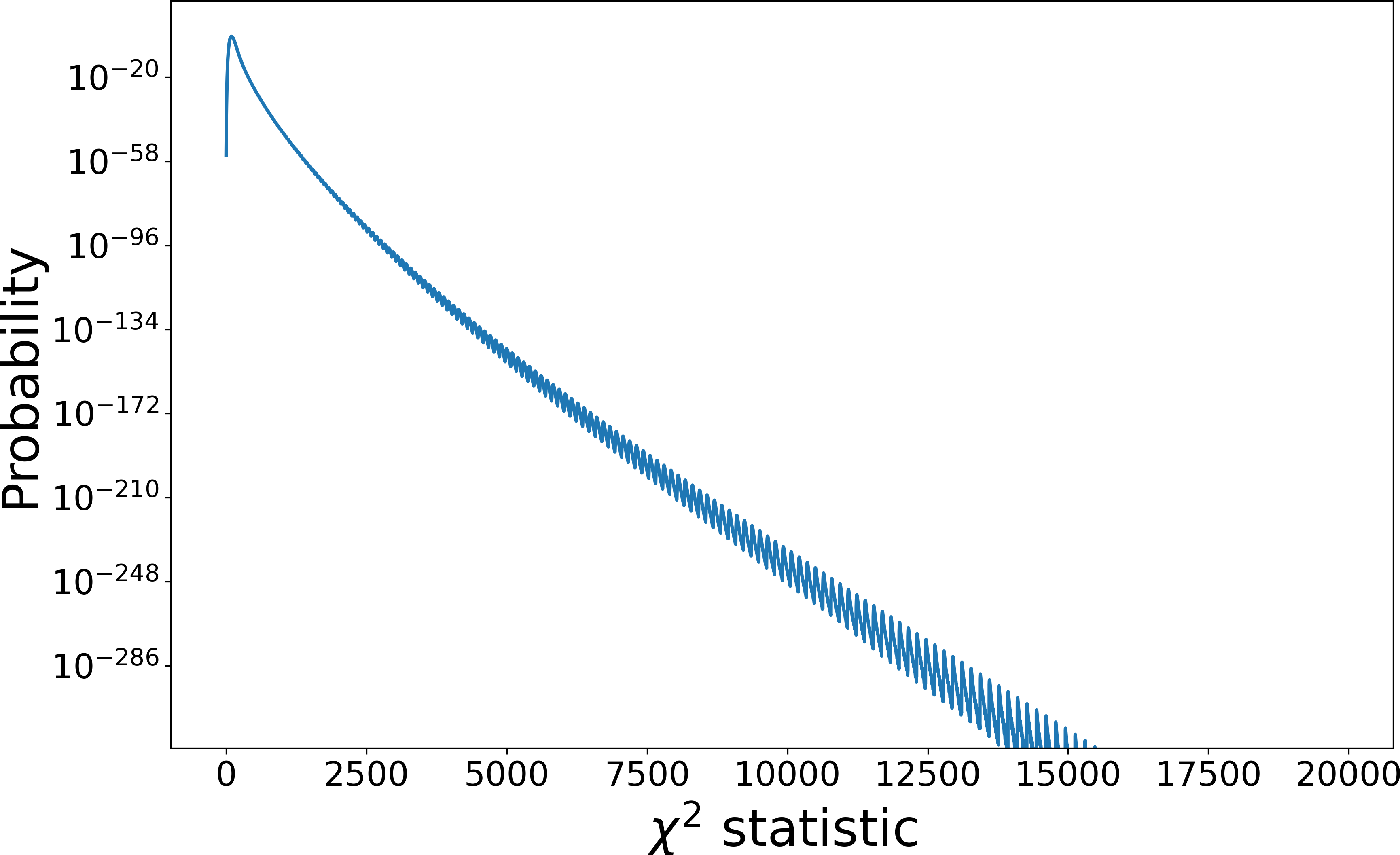}
	\label{fig:log_p_N_200_n_100}
	}%
	\subfloat[]{
	\includegraphics[width=0.45\linewidth]{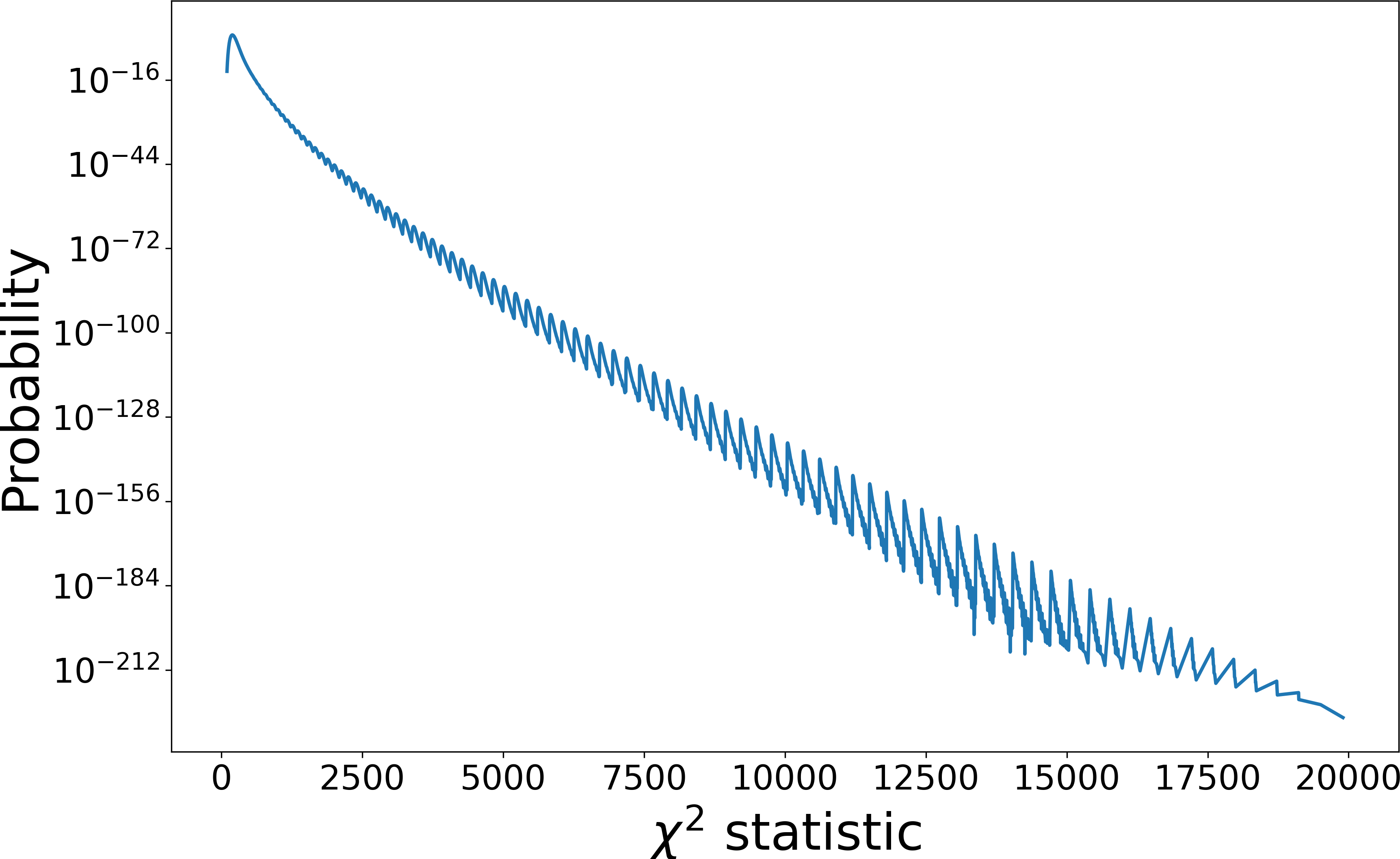}
	\label{fig:log_p_N_100_n_200}
	}\\

    \caption{Probabilities of the exact distribution of the chi-squared statistic for uniform histograms, plotted on a logarithmic scale on the y-axis, for the following cases: a)~$N=10$ and $n=20$, b)~$N=55$ and $n=10$, c)~$N=200$ and $n=100$, and d)~$N=100$ and $n=200$. The tail of the distribution often exhibits a log-linear trend with a noticeable zig-zag shape.}
	\label{fig:log_p}
    
\end{figure*}

\subsection{Implementation}
\label{subsec:algorithm}

The procedure in Section~\ref{subsec:specifics} is implemented in Algorithm~\ref{alg:p}. Line 12 skips calculations when the previous count is zero, significantly improving performance. This is possible by reversing the calculation direction compared to Section~\ref{subsubsec:added}: instead of computing a current count from all required previous counts, Algorithm~\ref{alg:p} adds a previous count to all current counts that require it. This allows skipping previous zero counts, 
reduces computation time, and slightly changes notation, e.g., by using $\binom{r}{nl}$ in Line 16, since $nl$ includes the last bin. In short, one calculation direction is simpler to notate, the other one is faster, but they both give the same final results.

Other implementation improvements are possible but are not detailed here, as they are not essential. For example, instead of iterating over all values from $0$ to $N^2$ in line 10 of Algorithm~\ref{alg:p}, one can pre-calculate all possible chi-squared values and iterate only over those, as demonstrated in the code available online. This approach yields slight performance improvements, but the theoretical complexity remains unchanged.

\subsection{Other statistics}
\label{subsec:other}

The described approach is not limited to the chi-squared statistic; it can be applied to any arbitrarily chosen statistic. This flexibility is particularly valuable in cases where a suitable approximation is unavailable and the exact distribution of a statistic would offer significant advantages. For instance, using the absolute value instead of the squared difference is one such alternative, which could further reduce computational complexity. While a detailed examination of these possibilities falls outside the scope of this paper, it represents a compelling direction for future research due to possible applications.

\subsection{Non-uniform distributions}
\label{subsec:nonuniform}

Calculating the exact distribution of the chi-squared statistic is straightforward for the uniform distribution because all of the $n^N$ possible assignment sequences are equally likely. However, for a non-uniform distribution, the probabilities of these sequences differ, which complicates the calculation. To handle this, one can modify the existing method by slightly adjusting Eq.~\eqref{eq:dp} to reflect the different probabilities, and by updating the scaling in Eq.~\eqref{eq:p} and Eq.~\eqref{eq:p_reused}. Although these changes are relatively minor, the underlying theory is more involved. For this reason, non-uniform distributions are not further considered here and are beyond the scope of this paper.

\begin{figure*}[htb]
    \centering
    
    \subfloat[]{
	\includegraphics[width=0.45\linewidth]{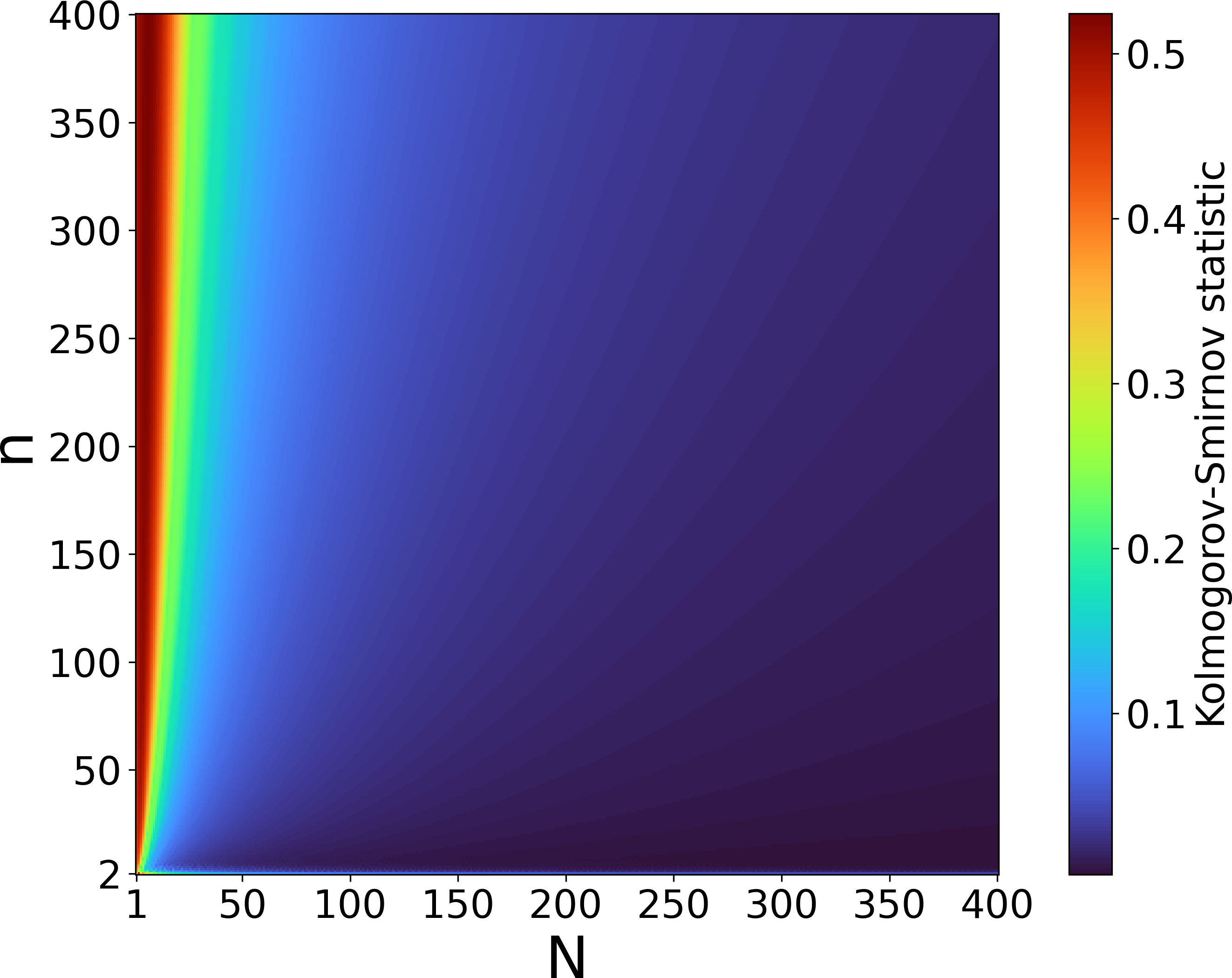}
	\label{fig:kss_1_2}
	}%
    \subfloat[]{
	\includegraphics[width=0.45\linewidth]{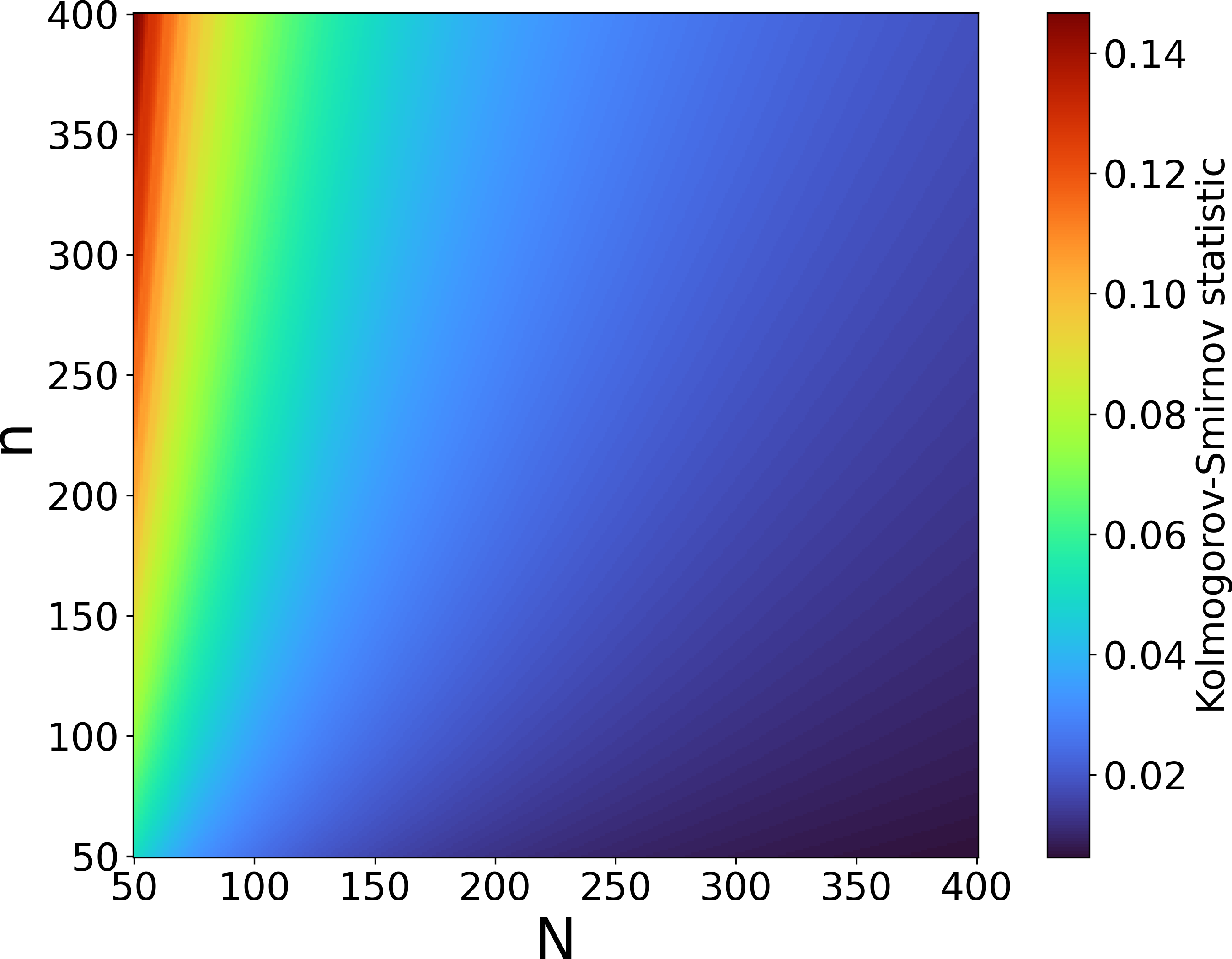}
	\label{fig:kss_50_50}
	}\\
	
    \label{fig:kss}
    
    \caption{The Kolmogorov-Smirnov statistic~\eqref{eq:kss} for various values or $N$ and $n$ with different starting values for the sake of better visualization in terms of value resolution: a)~$N=1$ and $n=2$, and b)~$N=50$ and $n=50$; lower is better.}
    
\end{figure*}

\subsection{Naming the whole process}
\label{subsec:name}

For simplicity, the entire process of computing the exact chi-squared distribution for discrete histograms using dynamic programming, culminating in Eq.\eqref{eq:p_reused} and Eq.\eqref{eq:p_discrete}, is referred to as the Zero-disparity Distribution Synthesis~(ZDS).

\section{Experimental results}
\label{sec:results}

All the results presented here can be repeated using the code given at~\url{https://github.com/DiscreteTotalVariation/ChiSquared}.

\subsection{Error measure}
\label{subsec:measure}

Since one of the goals is analyzing the approximation error of using the chi-squared distribution instead of the exact distribution, an appropriate metric for measuring this error can be the Kolmogorov-Smirnov~(K-S) statistic defined here as
\begin{equation}
    \label{eq:kss}
    D_{N, n} = \sup_{s} \left| P_{N, n}(s) - F_{n-1}(s) \right|.
\end{equation}
where $F_{n-1}(s)$ is the simplified notation for the CDF of the chi-squared distribution for $n-1$ degrees of freedom.

Examples comparing the exact distribution of the chi-squared statistic for uniform histograms to the chi-squared distribution approximation, along with the corresponding Kolmogorov–Smirnov (K-S) statistic, are shown in Fig.~\ref{fig:comparing} for various values of $N$ and $n$. The distribution tails are omitted because they appear flat, reflecting the negligible probability contribution of higher chi-squared values, as shown in Fig.~\ref{fig:log_p}.

\begin{figure}[htb]
    \centering
    
	\includegraphics[width=\linewidth]{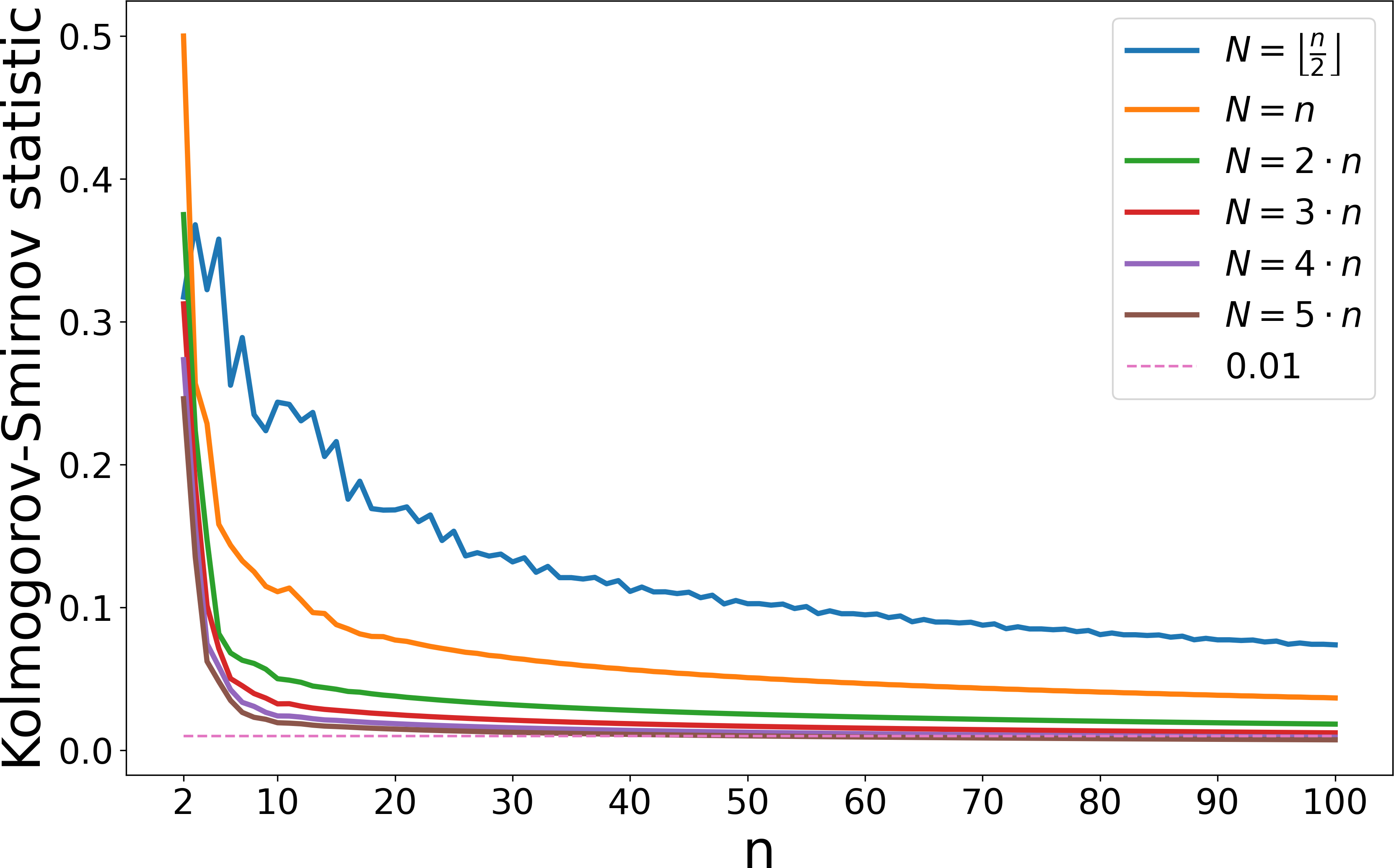}
	
    \caption{The Kolmogorov-Smirnov statistic~\eqref{eq:kss} for various values of $N$ and $n$ with fixed ratios; lower is better.}
	\label{fig:ratios}
    
\end{figure}

\subsection{Interpreting the error measure}
\label{subsec:interpreting}

Since the K-S statistic here compares two fully known distributions, there is no sampling variability, and its interpretation differs slightly. Its behavior can be evaluated using the normal approximation to the binomial distribution, which is considered acceptable when $Np \geq 10$ and $N\left(p-1\right) \geq 10$~\cite{moore2009introduction} where $N$ is the sample size and $p$ is the probability of success. For $p=0.5$ and $N=20$, the K-S statistic is approximately $0.001391$; for $p=0.01$ and $p=0.99$, it exceeds $0.02$. Across $p \in \{0.001, 0.002, \dots, 0.999\}$, the mean K-S value is about $0.011$. Despite variability, these results offer a rough baseline for acceptable approximation and may prompt further inquiry when considered alongside other findings that are shown here.

\begin{figure}[htb]
    \centering
    
	\includegraphics[width=\linewidth]{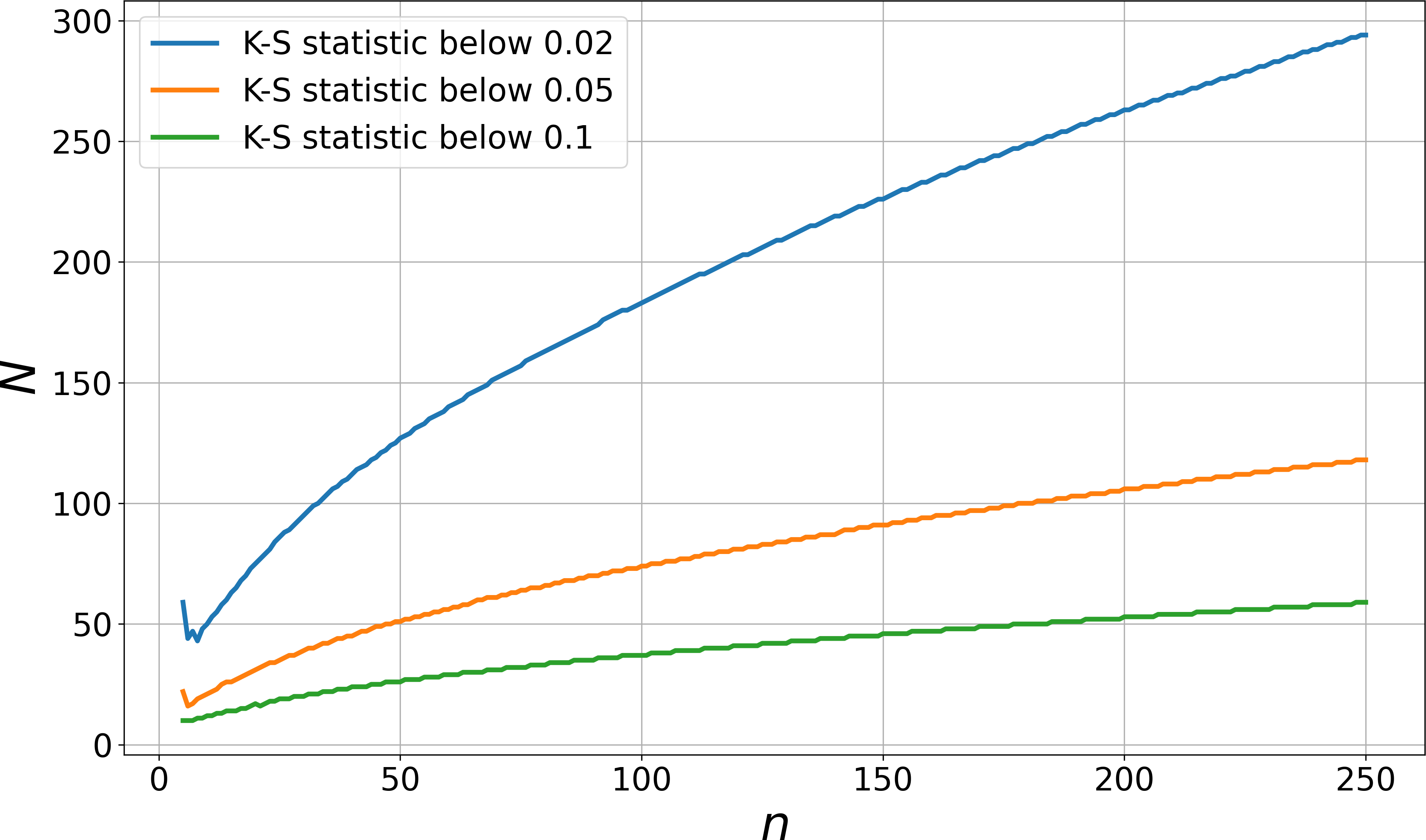}
	
    \caption{The lowest values of $N$ the for a given $n$ go below a Kolmogorov-Smirnov statistic threshold; lower is better.}
	\label{fig:thresholds}
    
\end{figure}

\begin{figure}[htb]
    \centering
    
	\includegraphics[width=\linewidth]{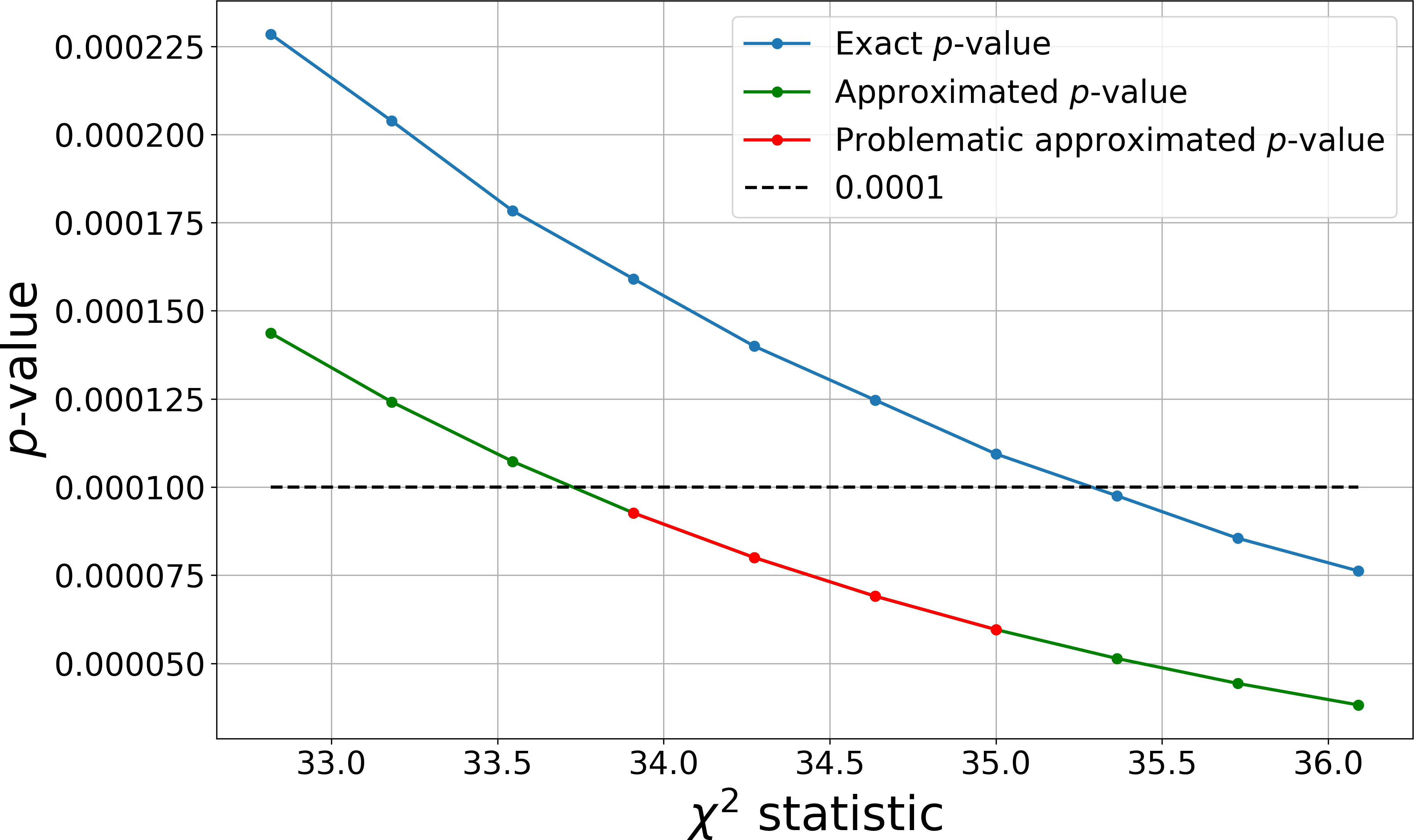}
	
    \caption{Exact and approximated $p$-values when $N=55$ and $n=10$ for various chi-squared statistics that achieve values close to the threshold of $10^{-4}$. The approximated $p$-values fall below the threshold earlier than the exact ones, thereby inflating type I error when relying on the approximation.}
	\label{fig:border}
    
\end{figure}

\subsection{The general rule of thumb}
\label{subsec:rule}

Fig.\ref{fig:ratios} shows that increasing the ratio $N/n$ reduces the K-S statistic. The only exception are the very small values of $n$. A common rule of thumb is that the chi-squared approximation is reliable when expected counts, which is $N/n$ here, are at least $5$~\cite{cochran1952chi2}. Fig.~\ref{fig:ratios} shows the K-S statistic behavior when expected counts are below, equal to, or above $5$. If $N<5\cdot n$, then the approximation is generally of lower quality. For small $n$, the expected count $5$ may be too low, while for large $n$, even the expected count $2$ may suffice. The K-S statistic is roughly $0.018$ for $N=200$ and $n=100$; it is roughly $0.062$ for $N=20$ and $n=4$. This indicates that the $N\geq 5\cdot n$ rule is suboptimal, and a more adaptive rule could be more useful.

As a matter of fact, a rarely mentioned confirmation of this can be found in some well cited sources~\cite{agresti2013categorical} where it is vaguely mentioned that if $n$ is large, then even a ratio of $N/n$ that equals $1$ is ''decent'' for the chi-squared approximation. For example, if the threshold of $0.02$ is used for the K-S statistic, then the lowest values of such $N$ and $n$ for which the K-S statistic falls below is $N=n=348$ with $0.019974$.

\subsection{Reconsidering the rule of thumb}
\label{subsec:reconsidered}

As noted in Section~\ref{subsec:interpreting}, an approximation seems valid if the K-S statistic is below, e.g., $0.02$. A practical rule of thumb should approximate this threshold as a function of $N$ and $n$. For a given $n$, Fig.~\ref{fig:thresholds} shows the lowest value of $N$ for which a given K-S statistic value lower than a given threshold is reached. It can be seen that the curves are sublinear and that even the slopes of their linear approximations are below $5$.

There is an additional point worth noting about Fig.~\ref{fig:thresholds}. The plot begins at $n=5$ because smaller values of $n$ correspond to significantly larger values of $N$ required to reach low K-S statistic values. For instance, when $n=2$, the K-S value drops below $0.02$ for the first time only when $N=1591$; for $n=3$, this occurs at $N=184$; and for $n=4$, at $N=77$. The phrase ''for the first time'' is used deliberately, as higher values of $N$ can result in the K-S statistic value increasing again. This is particularly evident for $n=3$, which exhibits notable oscillations. These irregularities at lower $n$ values stem from a violation of the independence assumption inherent to the chi-squared distribution. When $n$ is small, the bins are more interdependent because they collectively form a histogram with a fixed total, which contradicts the assumption of independent variables. This issue diminishes as $n$ increases, leading to more stable behavior with the K-S value falling.

\begin{figure*}[htb]
    \centering
    
    \subfloat[]{
	\includegraphics[width=0.45\linewidth]{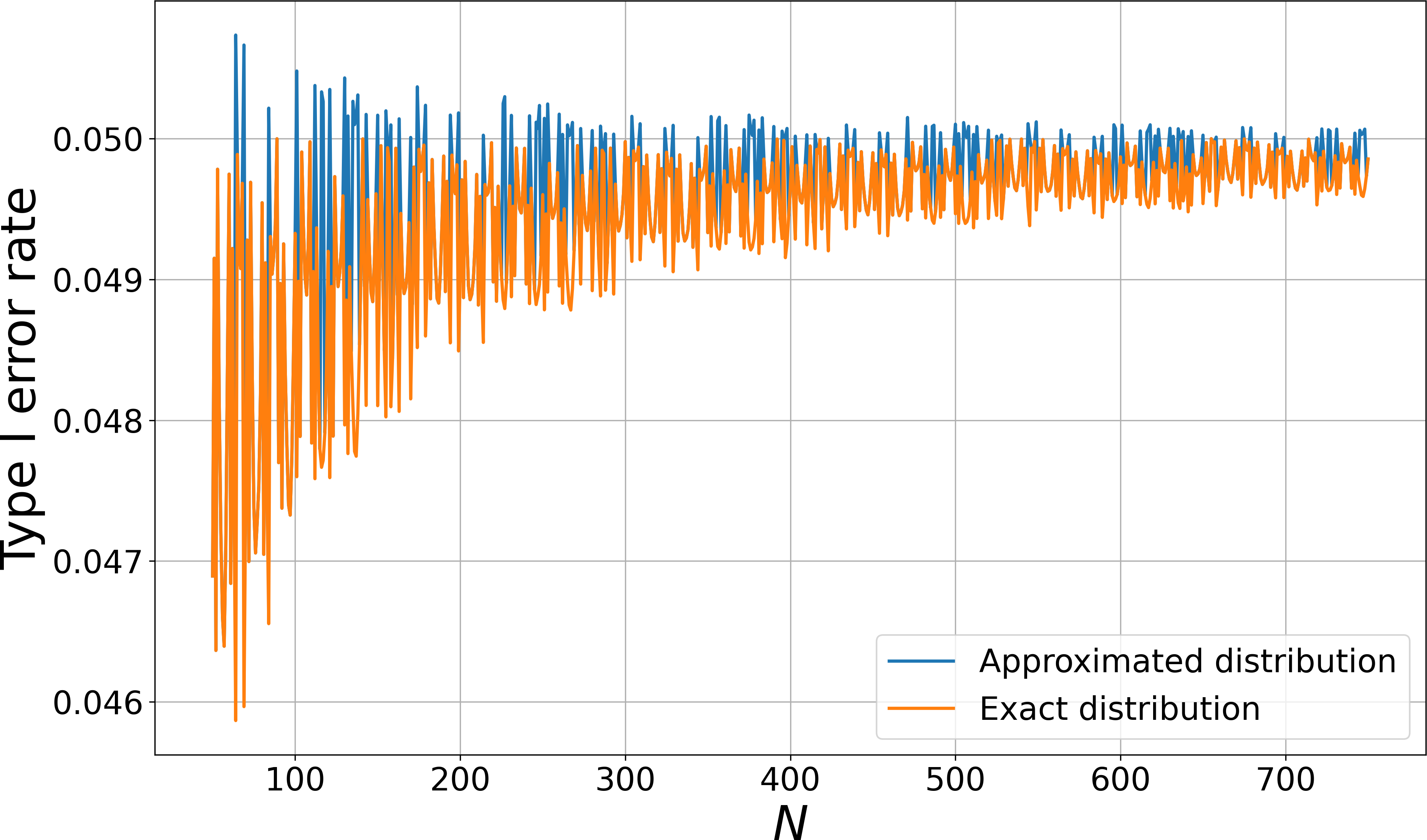}
	\label{fig:type_i_error_for_approximation_n_10_a_0.050000}
	}%
    \subfloat[]{
	\includegraphics[width=0.45\linewidth]{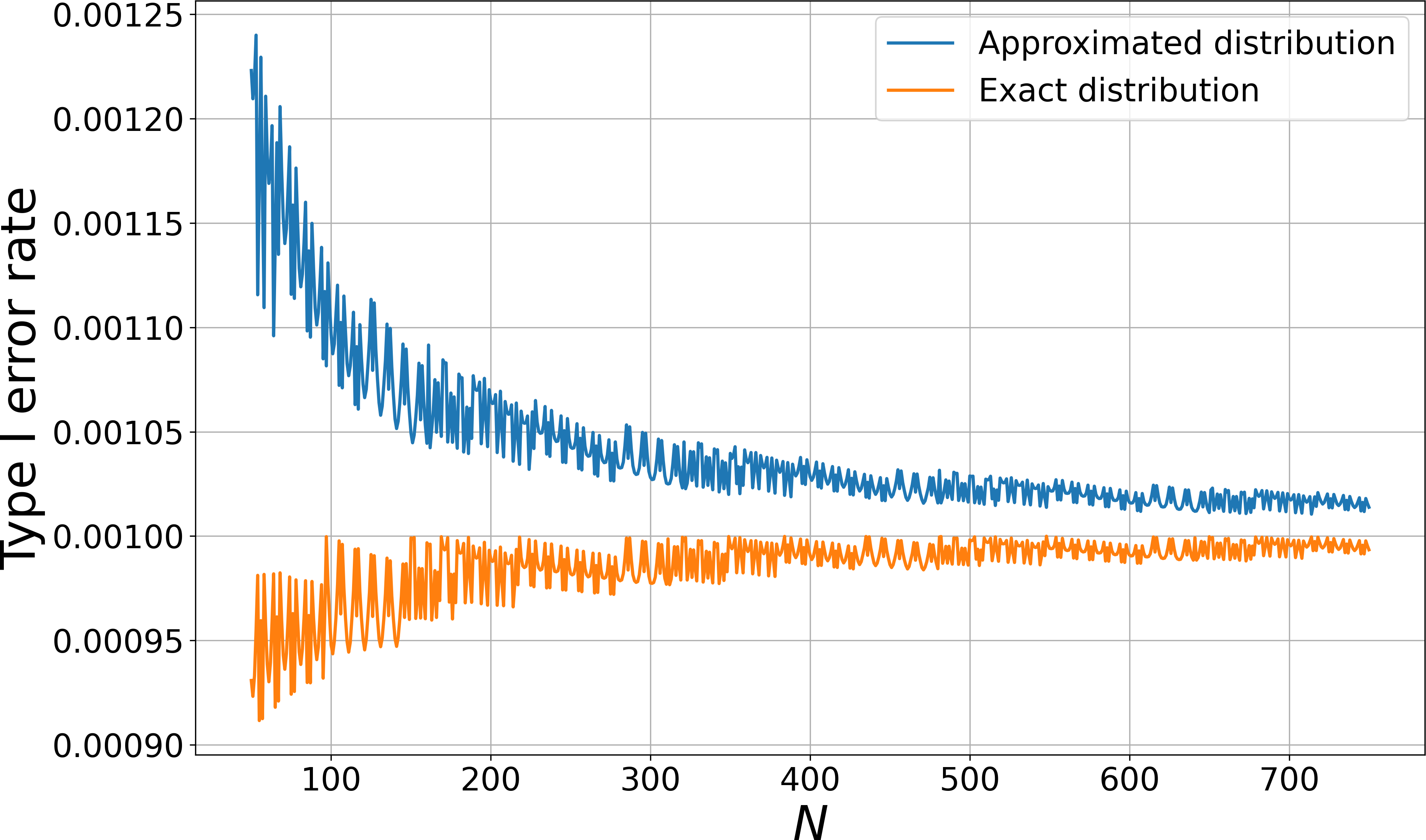}
	\label{fig:type_i_error_for_approximation_n_10_a_0.001000}
	}\\

	\subfloat[]{
	\includegraphics[width=0.45\linewidth]{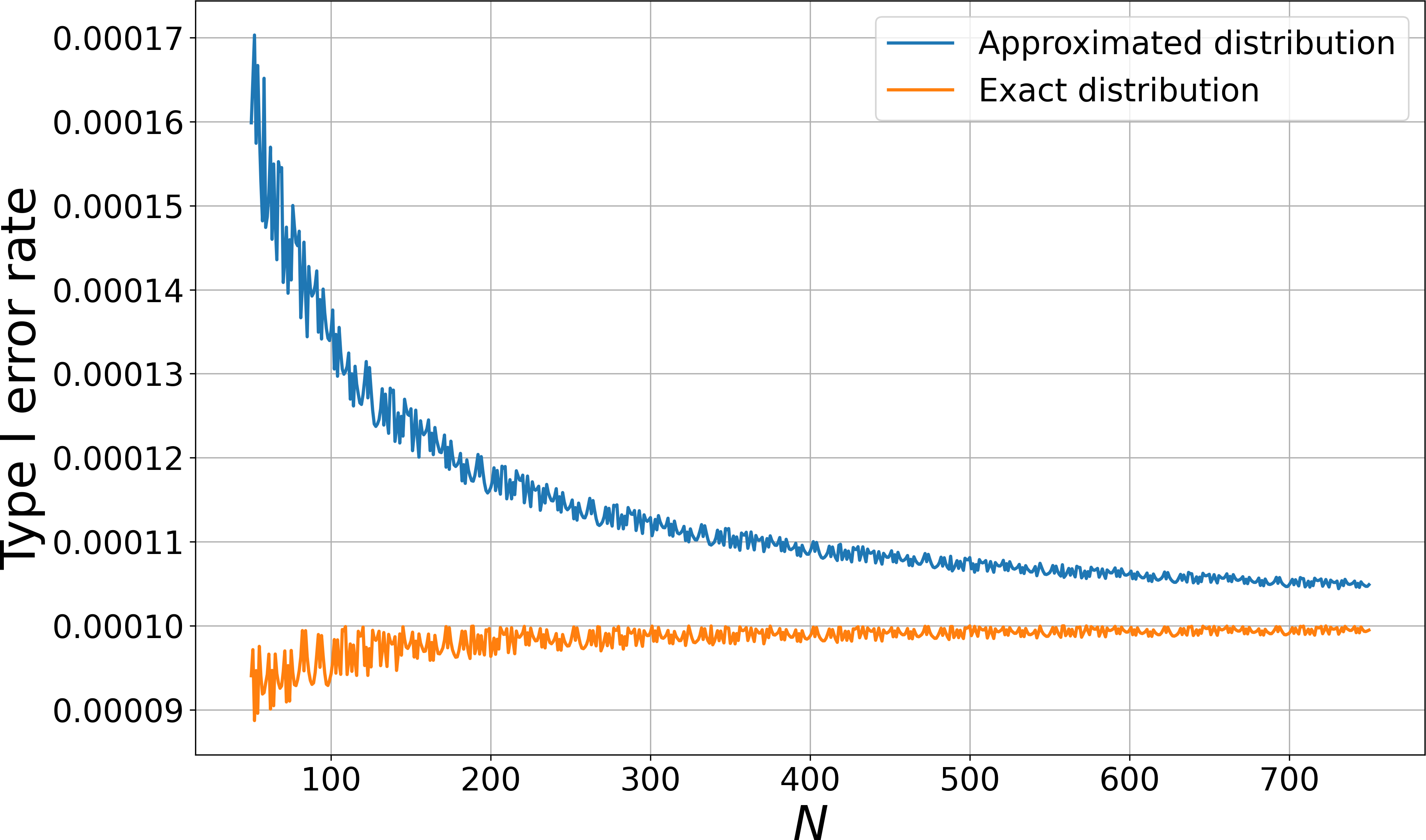}
	\label{fig:type_i_error_for_approximation_n_10_a_0.000100}
	}%
	\subfloat[]{
	\includegraphics[width=0.45\linewidth]{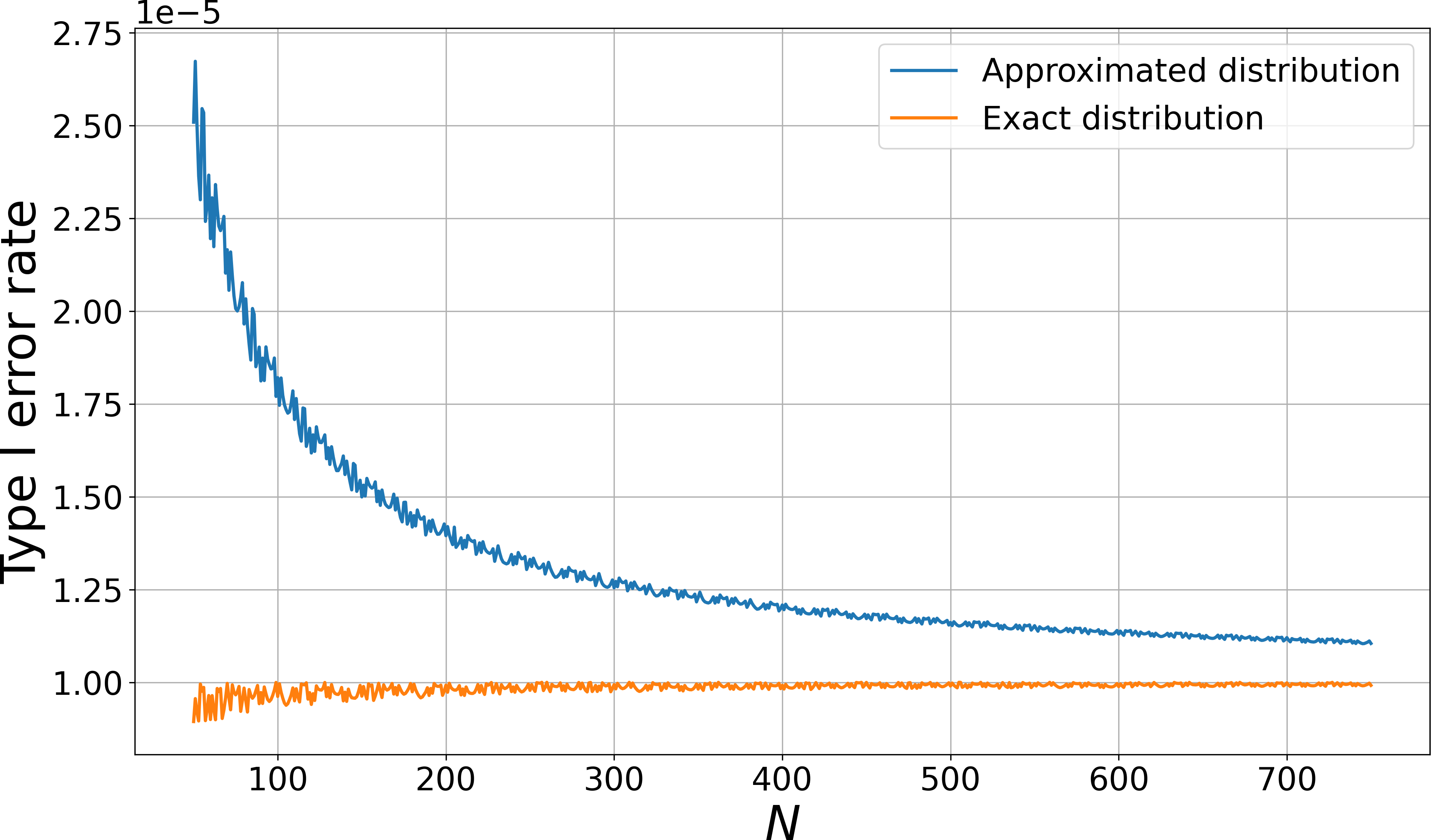}
	\label{fig:type_i_error_for_approximation_n_10_a_0.000010}
	}\\

    \caption{The type I error rate for various values of $N$ when using Eq.~\ref{eq:p_continuous} and Eq.~\eqref{eq:p_discrete} with the following thresholds: a)~$\alpha=0.05$, b)~$\alpha=0.001$, c)~$\alpha=0.0001$, and d)~$\alpha=0.00001$. As $\alpha$ lowers, the type I error rate worsens for the common approximation.}
	\label{fig:type_i_error_for_approximation}
    
\end{figure*}

\subsection{The NIST example}
\label{subsec:nist}

The revised version of the statistical test suite for random and pseudorandom number generators for cryptographic applications~\cite{bassham2010sp} published by the National Institute of Standards and Technology~(NIST) describes among other things also the strategy for the statistical analysis of a random number generator~(RNG) consisting of five stages. The first three stages involve selecting an RNG, generating random binary sequences, and performing repeated statistical tests. In the fourth stage, the resulting $p$-values are assessed for uniformity using a chi-squared test on $n=10$ bins. A significance level, i.e., $p$-value threshold is $\alpha=0.0001$, and at least $N=55$ $p$-values are said to be required for statistical significance.

Using Eq.~\eqref{eq:p_continuous}, four chi-squared values yield $p$-values below the significance level $\alpha$, while the correct values from Eq.~\eqref{eq:p_discrete} lie above $\alpha$. Specifically, for chi-squared statistics near $33.91$, $34.27$, $34.63$, and $35$, the exact $p$-values are approximately $1.59 \times 10^{-4}$, $1.4 \times 10^{-4}$, $1.25 \times 10^{-4}$, and $1.09 \times 10^{-4}$, respectively. In contrast, the chi-squared approximation yields lower values: $9.26 \times 10^{-5}$, $8 \times 10^{-5}$, $6.91 \times 10^{-5}$, and $5.96 \times 10^{-5}$. This discrepancy stems from the overestimated tail probabilities under the continuous approximation. As shown in Fig.~\ref{fig:log_p_N_55_n_10}, the exact tail probabilities decrease almost logarithmically, which leads to a slower decline in exact $p$-values. Consequently, the approximated $p$-value falls too fast under the given $\alpha$ of $10^{-4}$ as shown in Fig.~\ref{fig:log_p_N_55_n_10}, which is further increased by the wrong calculation described in Section~\ref{subsec:correct}. As a result, the type I error rate increases by over 50\%, from approximately $9.76 \times 10^{-5}$ to $1.59 \times 10^{-4}$. Fig.~\ref{fig:type_i_error_for_approximation_n_10_a_0.000100} shows how this persists for larger $N$ under fixed $n$ and $\alpha$, thus emphasizing the need for exact $p$-values in sensitive analyses. As further shown in Fig.~\ref{fig:type_i_error_for_approximation}, the problem is not immediately apparent for the commonly used $\alpha=0.05$, but it becomes significantly more pronounced if lower values of $\alpha$ are used.

\section{Conclusions}
\label{sec:conclusions}

A fast exact calculation of the chi-squared statistic distribution for discrete uniform histograms has been described. Relying on the newly available exact results, an analysis of the chi-squared distribution approximation was also given with some interesting and useful insights. The source code for using the proposed method and reproducing the presented results is made publicly available. Future work will include using dynamic programming for exact distribution calculations for histograms with non-uniform distributions, improving the chi-squared distribution approximation, redefining the currently used rule of thumb for choosing the relation between $N$ and $n$, and using another statistic instead of the chi-squared statistic.

\section{Acknowledgment}
\label{sec:acknowledgment}

The authors would like to thank Herman Zvonimir Do{\v{s}}ilovi{\'{c}} for providing the infrastructure used in part of the experiments.

\balance
\bibliographystyle{IEEEtran}
\bibliography{literature}

\begin{thebibliography}{10}
\providecommand{\url}[1]{#1}
\csname url@samestyle\endcsname
\providecommand{\newblock}{\relax}
\providecommand{\bibinfo}[2]{#2}
\providecommand{\BIBentrySTDinterwordspacing}{\spaceskip=0pt\relax}
\providecommand{\BIBentryALTinterwordstretchfactor}{4}
\providecommand{\BIBentryALTinterwordspacing}{\spaceskip=\fontdimen2\font plus
\BIBentryALTinterwordstretchfactor\fontdimen3\font minus
  \fontdimen4\font\relax}
\providecommand{\BIBforeignlanguage}[2]{{%
\expandafter\ifx\csname l@#1\endcsname\relax
\typeout{** WARNING: IEEEtran.bst: No hyphenation pattern has been}%
\typeout{** loaded for the language `#1'. Using the pattern for}%
\typeout{** the default language instead.}%
\else
\language=\csname l@#1\endcsname
\fi
#2}}
\providecommand{\BIBdecl}{\relax}
\BIBdecl

\bibitem{knuth1997art}
D.~E. Knuth, \emph{The Art of Computer Programming, Volume 2: Seminumerical
  Algorithms}.\hskip 1em plus 0.5em minus 0.4em\relax Addison-Wesley, 1997.

\bibitem{bassham2010sp}
L.~E. Bassham~III, A.~L. Rukhin, J.~Soto, J.~R. Nechvatal, M.~E. Smid, E.~B.
  Barker, S.~D. Leigh, M.~Levenson, M.~Vangel, D.~L. Banks \emph{et~al.},
  ``{S}{P} 800-22 {R}evision 1a. {A} {S}tatistical {T}est {S}uite for {R}andom
  and {P}seudorandom {N}umber {G}enerators for {C}ryptographic
  {A}pplications,'' 2010.

\bibitem{lipshitz1991quantization}
S.~P. Lipshitz, R.~A. Wannamaker, and J.~Vanderkooy, ``{Q}uantization and
  {D}ither: {A} {T}heoretical {S}urvey,'' in \emph{Audio Engineering Society
  Convention 91}.\hskip 1em plus 0.5em minus 0.4em\relax Audio Engineering
  Society, 1991.

\bibitem{kobus2024gaussian}
S.~Kobus, L.~Theis, and D.~G{\"u}nd{\"u}z, ``{G}aussian {C}hannel {S}imulation
  with {R}otated {D}ithered {Q}uantization,'' in \emph{2024 IEEE International
  Symposium on Information Theory (ISIT)}.\hskip 1em plus 0.5em minus
  0.4em\relax IEEE, 2024, pp. 1907--1912.

\bibitem{li2017probability}
X.~R. Li, \emph{{P}robability, {R}andom {S}ignals, and {S}tatistics}.\hskip 1em
  plus 0.5em minus 0.4em\relax CRC press, 2017.

\bibitem{bland2013baseline}
M.~Bland, ``{D}o {B}aseline p-values {F}ollow a {U}niform {D}istribution in
  {R}andomised {T}rials?'' \emph{PloS one}, vol.~8, no.~10, p. e76010, 2013.

\bibitem{bailey1995statistical}
N.~T. Bailey, \emph{Statistical methods in biology}.\hskip 1em plus 0.5em minus
  0.4em\relax Cambridge university press, 1995.

\bibitem{montgomery2020introduction}
D.~C. Montgomery, \emph{Introduction to statistical quality control}.\hskip 1em
  plus 0.5em minus 0.4em\relax John wiley \& sons, 2020.

\bibitem{vonta2008statistical}
F.~Vonta, M.~Nikulin, N.~Limnios, and C.~Huber-Carol, \emph{{S}tatistical
  {M}odels and {M}ethods for {B}iomedical and {T}echnical {S}ystems}.\hskip 1em
  plus 0.5em minus 0.4em\relax Springer Science \& Business Media, 2008.

\bibitem{davis2016handbook}
R.~A. Davis, S.~H. Holan, R.~Lund, and N.~Ravishanker, \emph{{H}andbook of
  {D}iscrete-{V}alued {T}ime {S}eries}.\hskip 1em plus 0.5em minus 0.4em\relax
  CRC Press, 2016.

\bibitem{dasgupta2008asymptotic}
A.~DasGupta, \emph{{A}symptotic {T}heory of {S}tatistics and
  {P}robability}.\hskip 1em plus 0.5em minus 0.4em\relax Springer, 2008, vol.
  180.

\bibitem{larntz1978small}
K.~Larntz, ``{S}mall {S}ample {C}omparisons of {E}xact {L}evels for
  {C}hi-{S}quare {G}oodness-of-{F}it {S}tatistics,'' \emph{Journal of the
  American Statistical Association}, vol.~73, no. 362, pp. 253--263, 1978.

\bibitem{cochran1952chi2}
W.~G. Cochran, ``{T}he $\chi$2 {T}est of {G}oodness of {F}it,'' \emph{The
  Annals of Mathematical Statistics}, pp. 315--345, 1952.

\bibitem{cochran1954some}
------, ``{S}ome {M}ethods for {S}trengthening the {C}ommon $\chi^2$ {T}ests,''
  \emph{Biometrics}, vol.~10, no.~4, pp. 417--451, 1954.

\bibitem{yarnold1970minimum}
J.~K. Yarnold, ``{T}he {M}inimum {E}xpectation in $\chi^2$ goodness of {F}it
  {T}ests and the {A}ccuracy of {A}pproximations for the {N}ull
  {D}istribution,'' \emph{Journal of the American Statistical Association},
  vol.~65, no. 330, pp. 864--886, 1970.

\bibitem{andres2000minimum}
A.~M. Andr{\'e}s and I.~H. Tejedor, ``On the minimum expected quantity for the
  validity of the chi-squared test in 2 $\times$ 2 tables,'' \emph{Journal of
  Applied Statistics}, vol.~27, no.~7, pp. 807--820, 2000.

\bibitem{decoursey2003statistics}
W.~DeCoursey, \emph{{S}tatistics and {P}robability for {E}ngineering
  {A}pplications}.\hskip 1em plus 0.5em minus 0.4em\relax Elsevier, 2003.

\bibitem{moore2009introduction}
D.~S. Moore, G.~P. McCabe, and B.~A. Craig, \emph{{I}ntroduction to the
  {P}ractice of {S}tatistics}.\hskip 1em plus 0.5em minus 0.4em\relax WH
  Freeman New York, 2009, vol.~4.

\bibitem{bluman2014elementary}
A.~Bluman, \emph{{E}lementary {S}tatistics: {A} {S}tep by {S}tep {A}pproach
  9e}.\hskip 1em plus 0.5em minus 0.4em\relax McGraw Hill, 2014.

\bibitem{siegel2016practical}
A.~F. Siegel, \emph{{P}ractical {B}usiness {S}tatistics}.\hskip 1em plus 0.5em
  minus 0.4em\relax Academic Press, 2016.

\bibitem{lock2020statistics}
R.~H. Lock, P.~F. Lock, K.~L. Morgan, E.~F. Lock, and D.~F. Lock,
  \emph{{S}tatistics: {U}nlocking the {P}ower of {D}ata}.\hskip 1em plus 0.5em
  minus 0.4em\relax John Wiley \& Sons, 2020.

\bibitem{agresti2013categorical}
A.~Agresti, \emph{{C}ategorical {D}ata {A}nalysis}.\hskip 1em plus 0.5em minus
  0.4em\relax John Wiley \& Sons, 2013.

\bibitem{yates1934contingency}
F.~Yates, ``Contingency tables involving small numbers and the $\chi^2$ test,''
  \emph{Supplement to the Journal of the Royal Statistical Society}, vol.~1,
  no.~2, pp. 217--235, 1934.

\bibitem{conover1974some}
W.~J. Conover, ``Some reasons for not using the yates continuity correction on
  2$\times$ 2 contingency tables,'' \emph{Journal of the American Statistical
  Association}, vol.~69, no. 346, pp. 374--376, 1974.

\bibitem{bellman2015applied}
R.~E. Bellman and S.~E. Dreyfus, \emph{{A}pplied {D}ynamic
  {P}rogramming}.\hskip 1em plus 0.5em minus 0.4em\relax Princeton university
  press, 2015, vol. 2050.

\bibitem{perkins2011chi2}
W.~Perkins, M.~Tygert, and R.~Ward, ``$\chi$2 and classical exact tests often
  wildly misreport significance; the remedy lies in computers,'' \emph{Uploaded
  to ArXiv}, 2011.

\bibitem{siegel1979noncentral}
A.~F. Siegel, ``The noncentral chi-squared distribution with zero degrees of
  freedom and testing for uniformity,'' \emph{Biometrika}, vol.~66, no.~2, pp.
  381--386, 1979.

\bibitem{bishop2007discrete}
Y.~M. Bishop, S.~E. Fienberg, and P.~W. Holland, \emph{{D}iscrete
  {M}ultivariate {A}nalysis: {T}heory and {P}ractice}.\hskip 1em plus 0.5em
  minus 0.4em\relax Springer Science \& Business Media, 2007.

\bibitem{indrayan2017medical}
A.~Indrayan and R.~K. Malhotra, \emph{{M}edical {B}iostatistics}.\hskip 1em
  plus 0.5em minus 0.4em\relax Chapman and Hall/CRC, 2017.

\bibitem{bonetti2019computing}
M.~Bonetti, P.~Cirillo, and A.~Ogay, ``Computing the exact distributions of
  some functions of the ordered multinomial counts: Maximum, minimum, range and
  sums of order statistics,'' \emph{Royal Society Open Science}, vol.~6,
  no.~10, p. 190198, 2019.

\bibitem{resin2023simple}
J.~Resin, ``{A} {S}imple {A}lgorithm for {E}xact {M}ultinomial {T}ests,''
  \emph{Journal of Computational and Graphical Statistics}, vol.~32, no.~2, pp.
  539--550, 2023.

\end{thebibliography}

\end{document}